\documentclass{article}

\usepackage{amssymb}
\usepackage[T1]{fontenc}
\usepackage[latin1]{inputenc}
\usepackage{amssymb}
\usepackage{amsmath}
\usepackage{graphicx}
\begin{document}

\begin{center}
\textbf{\bigskip }

\textbf{Characterization of Failure Mechanism in Composite Materials Through
Fractal Analysis of Acoustic Emission Signals}$^{\S }$

\bigskip

F. E. Silva$^{a}$, L. L. Gon\c{c}alves$^{a,b\dag },$ D. B. B. Fereira$^{c}$
and J. M. A. Rebello$^{d}$

$^{a}$Programa de Mestrado em Engenharia e Ci\^{e}ncia de Materiais

Universidade Federal do Cear\'{a},

Campus do Pici, Bloco 714, 60455-760 Fortaleza, CE, Brazil

$^{b}$Departamento de Fisica Geral, Universidade de S\~{a}o Paulo,

C.P. 66318, 05315-970, S\~{a}o Paulo, SP, Brazil

$^{c}$Funda\c{c}\~{a}o de Apoio \`{a} Escola T\'{e}cnica do Rio de Janeiro

21311- 280 Rio de Janeiro, RJ, Brazil

$^{d}$Departmento de Engenharia Metal\'{u}rgica e de Materiais

Universidade Federal do Rio de Janeiro

CP 68505, 21945-970 Rio de Janeiro, RJ, Brazil

\bigskip

Abstract
\end{center}

In this paper it is presented a detailed numerical investigation of acoustic
emission signals obtained from test samples of fibreglass reinforced
polymeric matrix composites, when subjected to tensile and flexural tests.\
Various fractal indices, characteristic of the signals emitted at the
different structural failures of the test samples and which satisfy
non-stationary distributions, have been determined. From the results
obtained for these indices, related to the Hurst analysis, detrended
fluctuation analysis, minimal cover analysis and to the boxcounting
dimension analysis, it has been shown they can discriminate the different
failure mechanisms and, threfore, they constitute their signature.

$\bigskip $

$\bigskip $

$\bigskip $

$\bigskip $

\noindent $^{\dagger }$Corresponding author.

\noindent E-mail: lindberg@fisica.ufc.br

\noindent On sabbatical leave from:

\noindent Departamento de Fisica

\noindent Universidade Federal do Cear\'{a}

\noindent Campus do Pici, Caixa Postal 6030

\noindent 60451-970 Fortaleza, Cear\'{a}, Brazil

$\bigskip $

$\bigskip $

\noindent $^{\S }$Work partially financed by the Brazilian agencies CNPq,
Finep (CT-Petro) and Capes.

\pagebreak

\textbf{1. INTRODUCTION}

In a recent paper Ferreira et al [1] discuss the characterization of failure
mechanisms that occur in fibreglass reinforced polymeric matrix composites
when subjected to tensile and flexural loads. The characterization was based
on the analysis of acoustic emission signals emitted by the composite during
the process of failure, which constitutes one the most important
non-destructive testing for the detection of structural flaws in composite
materials [2-5].

The samples studied were manufactured with E-glass fibre roving reinforced
DER 331 epoxy resin and its preparation and experimental conditions are
described in detail in ref. [1]. Besides tensile tests, flexural tests at
three- and four-points were also applied and four failure modes have been
observed, namely, matrix cracking, fibre braking, fibre/matrix \ debonding
and delamination.

The main purpose of the study was to find the signature of these failure
modes in the acustic emission signals. In order to identify these
signatures, the signals were studied by using Fourier spectral analysis and
wavelet analysis. Although relevant information has been obtained from these
analyses, the authors in ref. [1] have not been able to characterize in a
clear way the various failure mechanisms.

Therefore, in this paper we readdress the problem by looking at \ some
fractal properties of the acoustic emission signals. In particular, we
obtain the fractal indices related to the Hurst analysis [6], detrended
fluctuation analysis [7], minimal cover analysis[8] and to the boxcounting
dimension analysis [9], which will be used to characterize the different
failure modes.

These types of analysis have been widely used in the study of random
nonstationary series ranging from seismic [10] and climate data, [11] to
wind speed [12] \ and financial data [13], and in the study of different
music genres [14]. Their use in the characterization of acoustic signal has
been introduced by Duta and Barat [15] in the analysis of ultrasonics
backscattered signals obtained in the study of \ single crystal and
polycrystalline materials. More recently, Matos et al. [16] have used this
approach to characterize the ultrasonics backscattered signals obtained in
the study of the cast iron with lamellar, vermicular and spheroidal
microstructures.

The study presented in this paper extends the above mentioned analyses to a
new type of acoustic signals, namely, the ones obtained in the acoustic
emission nondestructive testing. The main objective of the work is to show
that the parameters determined from these analyses can characterize the
failure mechanisms in composite studied. To this aim and in order to
establish the parameters to be calculated, we present in section 2 a brief
review of the numerical analysis used in the treatment of the data, and \ in
section 3 we present and discuss the results obtained.

\bigskip \pagebreak

\textbf{\bigskip 2. NUMERICAL ANALYSIS}

The numerical treatment of the signals will be performed on data from
A-scan, which contains the amplitude of the acoustic emission signals as a
function of time. The parameters to be determined, as pointed out in the
introduction, will be obtained from the Hurst analysis (R/S analysis) [6],
detrended fluctuation analysis (DFA analysis) [7], minimal cover analysis
[8] and boxcounting analysis [9].

In order to make the paper self-contained and to introduce the notation, we
will \ present a brief review of the these numerical techniques which will
be used in the analysis of the temporal series. They will be identified as
the set of random values $\{y_{i}\}$, where the label $i$ corresponds to the
time variable, which satisfy nonstationary distributions.

\qquad \textbf{2.1 Hurst analysis}

The R/S analysis will provide information on the temporal correlations, on
various time-scales, of the data. Given the temporal series $\{y_{i}\},$
with $N$ terms ($1\leqslant i\leqslant N)$, we define the average in the
interval $n$ as

\begin{equation}
<y>_{n}=\frac{1}{n}\sum_{i=1}^{n}y_{i},  \
\end{equation}%
$\bigskip $

and the accumulated deviation from the mean as

\bigskip

\begin{equation}
Y(j,n)=\sum_{i=1}^{j}\left( y_{i}-<y>_{n}\right) ,  \
\end{equation}

where $n$ varies from $2$ to $N$.

From these results, we can also define in the interval $n$ the range $R(n)$
of the accumulated deviation in the form

\bigskip

\begin{equation}
R(n)=\max_{1\leqslant j\leqslant n}Y(j,n)-\min_{1\leqslant
j\leqslant n}Y(j,n),  \
\end{equation}%
and the standard deviation $S(n)$ as

\begin{equation}
S(n)=\sqrt{\frac{\sum_{j}^{n}\left( y_{j}-<y>_{n}\right) }{n}.}  \
\end{equation}

Finally, we can obtain the rescaled range $R(n)/S(n)$which should satisfy
the scaling relation

\bigskip

\begin{equation}
\frac{R(n)}{S(n)}\thicksim n^{H},  \
\end{equation}

where $H$ is the Hurst exponent [6].

\qquad In the scaling regime, the previous expression can be written as

\begin{equation}
\frac{R(n)}{S(n)}=A_{H}n^{H},  \
\end{equation}%
\bigskip

\noindent which defines the amplitude $A_{H}.$ Although this parameter has
no universal characteristic, as the amplitudes in the scaling laws in
critical phenomena [17] where they are related to the interactions, they can
be used however as an additional parameter to characterize the temporal
series.

\qquad \textbf{2.2 Detrended fluctuation analysis}

The DFA analysis [7] aims to study the temporal correlations by eliminating
the spurious trends in the data which can conduct to misleading results. The
method consists initially in obtaining a new integrated temporal series $%
\{z_{i}\},$ from the original one $\{y_{i}\},$ given by

\bigskip

\begin{equation}
z_{j}=\sum_{i=1}^{j}\left( y_{i}-<y>\right) ,  \
\end{equation}

where the average $<y>$ is defined as

\bigskip

\begin{equation}
<y>=\frac{1}{N}\sum_{i=1}^{N}y_{i}.  \
\end{equation}

\bigskip

In the following step the series is divided in time intervals of width $n,$
and an order-$l$ polynomial  is fitted in each interval, and we identify the
analysis as DFA-$l$. Then, the detrended variation function of order $l$ in
the interval $j$, $\Delta _{j}^{l}(j),$ is obtained by subtracting the local
trend contained in the fitted polynomial, and is given by

\begin{equation}
\Delta _{j}^{l}(n)=\sum_{i=(j-1)n+1}^{jn}\left(
z_{i}-z_{i}^{l}\right) ^{2}, \
\end{equation}

where $z_{i}^{l}$ is the value from the fitted polynomial.

Finally, we calculate the mean root square fluctuation $F^{l}(n)$

\bigskip

\begin{equation}
F^{l}(n)=\sqrt{\frac{1}{N}\sum_{j=1}^{int[N/n]}\Delta _{j}^{l}(n),}
\
\end{equation}%
\bigskip

\noindent which should scale as

\bigskip

\begin{equation}
F^{l}(n)\thicksim n^{\alpha },  \
\end{equation}

\bigskip

\noindent where $\alpha $ is the scaling exponent.

The detrended fluctuation analysis that we will present will be restricted
to the linear case, namely, DFA-1. As in the case of the R/S analysis,
eq.(11) can be written in the scaling regime as

\begin{equation}
F^{l}(n)=A_{\alpha }n^{\alpha },  \
\end{equation}%
\bigskip

which also defines a new characteristic parameter $A_{\alpha }.$

\qquad \textbf{2.3 Minimal cover analysis}

This method has been recently introduced [8], and it relates the minimal
area necessary to cover a given plane curve, in a specified scale, to a
power law behaviour. The scale is introduced by dividing the domain of
definiton of the function in $n$ intervals of width $\delta .$ In each
interval $j$ $(1\leqslant j\leqslant n)$ we can associate a rectangle of
base $\delta $ and height $A(j)$ defined as

\bigskip

\begin{equation}
A_{j}=\max \{y_{i},i\epsilon \lbrack j,j+\delta ]\}-\min
\{y_{i},i\epsilon \lbrack j,j+\delta ]\},  \
\end{equation}%
\bigskip

such that the minimal area will be given by

\bigskip

\begin{equation}
S(\delta )=\sum_{j=1}^{n}A_{j}\delta .  \
\end{equation}

In the scaling region, $S(\delta )$ should behave as

\bigskip

\begin{equation}
S(\delta )\thicksim \delta ^{2-D_{\mu }},  \
\end{equation}

\bigskip

where $D_{\mu }$ is the minimal cover dimension, which is equal to 1 when
the curve presents no fractality. We can also define a new exponent $\mu $
given by

\begin{equation}
\mu =D_{\mu }-1,  \
\end{equation}

which measures the fractality of the curve and satisfies the scaling relation

\begin{equation}
V(\delta )\thicksim \delta ^{-\mu },  \
\end{equation}

\bigskip where $V(\delta )$ is the summmation of the heights of the
rectangles

\begin{equation}
V(\delta )=\sum_{j=1}^{n}A_{j}.  \
\end{equation}

The amplitude $A_{\mu },$ as in the previous cases, is defined in the
expression

\bigskip

\begin{equation}
V(\delta )=A_{\mu }\delta ^{-\mu },  \
\end{equation}

\bigskip

and it also constitutes a new characteristic parameter.

\bigskip \qquad \textbf{2.4 \ Boxcounting analysis}

The boxcounting dimension, which is one of the best known fractal dimension
[9], is easily defined and obtained numerically. It can be introduced in a
general d-dimensional euclidean space, where a hyper-volume is embedded, by
considering the number of hypercubes of side length $\delta ,$\textsl{\ }$%
\mathcal{N}$($\delta ),$ necessary to cover the entire volume. As $\delta
\rightarrow 0,$ $\mathcal{N}$($\delta )$ satisfies the scaling relation

\bigskip

\begin{equation}
\mathcal{N}(\delta )\thicksim \delta ^{D_{B}},  \
\end{equation}

\bigskip

where $D_{B}$ is the boxcounting fractal dimension.

For non-fractal objects, this dimension corresponds to the topological
dimension and, in particular, for continuous planar curves $D_{B}$ is equal
to 1.

The amplitude $A_{B}$ of the scaling relation is, in this case,
given by

\begin{equation}
\mathcal{N}(\delta )=A_{B}\delta ^{D_{B}},  \
\end{equation}

and it also constitutes a new characteristic parameter.

\bigskip \pagebreak

\textbf{3. RESULTS AND DISCUSSIONS}

A detailed description of the experimental setup and samples used for the
acquisition of the acoustic emission signal from the different tests is
presented by Ferreira et al. [1]. Besides the tensile test, the samples were
submitted to three- and four-point flexural tests. They have identified four
basic failure modes, namely, matrix cracking, fibre breaking, fibre/matrix
debonding and delamination. In the different tests, the failure mechanisms
were a result of a combination of these failure modes, and they are
presented in Table 1. We also present in this Table the acronyms for the
different specimens.

In order to reduce the noise in the data, the signals have been processed
with an adjacent low-pass filter with five points. For each type of specimen
the tests were carried out in \ 03 samples, which correspond to the number
of signals available for each kind of mechanical failure.

In Figs. 1-4 we present the various analyses made in a given signal from the
TEM specimen..These analyses are representative of the results obtained in
the study of the other signals. In the Hurst analysis, Fig. 1, the crossover
from short- to long- time correlations is always present. As can be seen in
Figs. 2-4, this crossover also exists in the box counting analysis and in
minimal cover analysis, but not in the DFA analysis. It should be noted that
the presence existence of this crossover on the fractal analysis is
characteristic of a multifractal behaviour.

Besides the eight parameters $H,A_{H},\alpha ,A_{\alpha },\mu ,A_{\mu
},D_{B},A_{B},$\ we can yet define an additional one which corresponds to
the standard deviation $\sigma $\ of the signal. This parameter has been
recently introduced in the context of the characterization of climate of
different regions in the United States from the analysis of maximum daily
temperature time series [11].

From what we have presented, this multi-dimensional space parameter
can be used to discriminate the various types of mechanical failure.
As it will be shown in the figures relating the characteristic
parameters, there is not a unique signature, since different
combinations of the indices can lead to the identification of the
signals.We have restricted our study to subspaces of the parameter
space, as they can provide the desired signature of the
signals. Explicitly we will consider the exponents $H,\alpha ,\mu $\ and $%
D_{B}$\ as functions of the standard deviation of the signal, $\sigma ,$\
and also as functions of the logarithm of its respective amplitude $A.$

These functions, which correspond to projections of the points of
the parameter space in different planes, are shown in Figs. 5-17.
Even considering that the data for each type of specimen consisted
of three samples only, which is a poor statistical sampling, we have
calculated the standard deviation of the variables presented as the
error bar on these figures. From the analysis of these figures we
can verify that the first discrimination attained is the separation
of the failures caused by traction from the ones caused by flexion.
This discrimination is clearly seen in the diagrams $H_{1}\times
\sigma ,$\ $H_{2}\times \sigma ,$\ $D_{B2}\times \sigma $\ and
$\alpha \times \sigma $\ which are presented in the Figs. 5, 6, 8
and 11, respectively. This separation is the easiest to be obtained
since the stress distributions in the samples are very different in
the two cases, and this has strong effect on the acoustic emission
signals.

By starting from any of these diagrams we can obtain the complete
discrimination of all the failures mechanisms and even distinguish the
results obtained from the three- and four-point flexural tests. This can be
achieved by using the complete set of figures, namely, Figs. 5-17. In terms
of these diagrams, all possible solutions of the problem are presented in
the tree type graph shown in Fig. 18. In the branches of the tree, we
designate the various failure modes and, in the nodes, we show the different
two-dimensional diagrams which can lead to the desired discrimination. As
can be seen on the solution tree, we can identify unmistakably all the
failure modes from the acoustic emission signals and, moreover, show that
there are multiple paths which lead to the identification we are looking for.

\bigskip \pagebreak

\textbf{REFERENCES}

[1] Ferreira DB, Silva RR da, Rebello JMA, Siqueira MHS. Failure

\ \ \ \ mechanism characterization in composite materials using spectral

\ \ \ \ analysis and the wavelet transform of acoustic emission\ signals.

\ \ \ \ \textit{Insight} 2004; 46: 282-9.

[2] McIntire P, editor.\textit{\ Nondestructive Testing Handbook- }

\ \ \ \ \textit{Acoustic Emission}, v. 5, 2nd edition. Columbus: ASNT; 1987.

[3] Wegman RF. \textit{Nondestructive Test Methods for Structural Composites.%
}

\textit{\ \ \ \ \ \ SAMPE Handbook Series}.Corvina: SAMPE; 1989.

[4] Summerscales J. NDT of advanced composites - an overview of the

\ \ \ \ \ \ possibilities, \textit{British Journal of Non-destructive
Testing }1990; 32:

\ \ \ \ \ 568-577.

[5] Clarke DJ. Thoughts on NDT for composites in the field. \textit{Insight}
1995;

\ \ \ \ \ 37: 938-939.

[6]. Hurst HE. Long-term capacity of reservoirs.\textit{\ Trans. Am. Soc.
Civ.}

\textit{\ \ \ \ \ \ Eng.} 1951; 116: 770-808.

[7]. Peng CK, Buldyrev V, Havlin S, Simmons M,\ Stanley HR, Goldberger

\ \ \ \ \ \ AL. Mosaic organization of DNA nucleotides.\textit{\ Phys. Rev.
E }1994; 49:

\ \ \ \ \ \ 1685-1689.

[8] Dubovikov MM, Starchenko NV, Dubovikov MS.\ Dimension of

\ \ \ \ \ minimal cover and fractal analysis of time series. \textit{Physica
A} 2004; 339:

\ \ \ \ \ \ 591-608.

[9] See e.g. Addison PS. \textit{Fractals and Chaos}. London: IOP; 1997.

[10] Telesca L, Lapenna V, Macchiato M. Mono- and multi-fractal

\ \ \ \ \ \ investigation of scaling properties in temporal patterns
sequences of

\ \ \ \ \ \ seismic sequences. \textit{Chaos, Solitons \& Fractals }2004;
19: 1-15.

[11] Kurnaz ML. Detrended fluctuation analysis as a statistical tool to

\ \ \ \ \ \ \ monitor the climate. \textit{J. Stat. Mech.: Theor. Exp}.
2004; P07009.

[12] Govindan RB, Kantz H. Long-term correlations and multifractality

\ \ \ \ \ \ \ in surface wind speed. \textit{Europhys. Lett. }2004; 68:
184-190.

[13] Carbone A, Castelli G, Stanley HE. Time-dependent Hurst exponent

\ \ \ \ \ \ \ in financial time series. \textit{Physica A} in press.

[14] Jennings HD, Ivanov P Ch, da Silva PC, Viswanathan GM. \textit{Physica A%
}

\ \ \ \ \ \ 2004; 336: 585-594.

[15] Dutta D, Barat P. \textit{J. Acoust. Soc. Am}. 1995; 98: 938-942.

[16] Matos JMO, de Moura EP, K\"{u}ger SE, Rebello JMA. \textit{Chaos, }

\ \ \ \ \ \ \textit{Solitons \& Fractals} 2004; 19: 55-60.

[17] See e.g. Plischke M, Bergensen B. \textit{Equilibrium Statistical
Physics}.

\ \ \ \ \ \ \ Singapore: World Scientific; 1994.

\bigskip \pagebreak

\textbf{Table 1. Failure modes and respective mechanical tests for the }

\textbf{different specimen types.}

\begin{tabular}{|c|c|c|}
\hline
\textbf{Test} & \textbf{Specimen type} & \textbf{Failure modes} \\ \hline
Tensile & Epoxy (TME*) & Matrix cracking \\ \hline
Tensile & Fibre/Epoxy(TLEV*) & $%
\begin{array}{c}
{Fibre breaking} \\
{Matrix cracking} \\
{Fibre/matrix debonding}%
\end{array}%
$ \\ \hline
Tensile & Fibre/Epoxy(TTEV*) & Matrix cracking \\ \hline
4-point flexural & Fibre/Epoxy(F41*) & $%
\begin{array}{c}
{Fibre breaking} \\
{Matrix cracking} \\
{Fibre/matrix debonding}%
\end{array}%
$ \\ \hline
4-point flexural & Fibre/Epoxy(F41S*) & $%
\begin{array}{c}
{Fibre breaking} \\
{Matrix cracking} \\
{Fibre/matrix debonding}%
\end{array}%
$ \\ \hline
4-point flexural & Fibre/Epoxy(F42*) & $%
\begin{array}{c}
{Fibrebreaking} \\
{Matrix cracking} \\
{Fibre/matrixde bonding} \\
{Delamination}%
\end{array}%
$ \\ \hline
3-point flexural & Fibre/Epoxy(F31*) & $%
\begin{array}{c}
{Fibre breaking} \\
{Matrix cracking} \\
{Fibre/matrix debonding}%
\end{array}%
$ \\ \hline
3-point flexural & Fibre/Epoxy(F31S*) & $%
\begin{array}{c}
{Fibre breaking} \\
{Matrix cracking} \\
{Fibre/matrix debonding}%
\end{array}%
$ \\ \hline
3-point flexural & Fibre/Epoxy(F41*) & $%
\begin{array}{c}
{Fibre breaking} \\
{Matrix cracking} \\
{Fibre/matrix debonding} \\
{Delamination}%
\end{array}%
$ \\ \hline
\multicolumn{3}{|l|}{{*}Specimen acronyms.} \\ \hline
\multicolumn{3}{|l|}{F41S and F31S identify the samples with surface
treatment of fibres} \\ \hline
\end{tabular}

\bigskip

\pagebreak

\textbf{Figure captions}

Fig. 1- Hurst analysis of a signal obtained from a TME specimen.

\ \ \ \ \ \ \ \ $H_{1}$ and $H_{2}$ are the Hurst exponents associated with

\ \ \ \ \ \ \ \ short- and long-time correlations, respectively.

Fig. 2- DFA analysis for the signal used in Fig. 1.

Fig. 3- Minimal cover analysis for the signal used in Fig. 1.

\ \ \ \ \ $\ \ \ \mu _{1}$\ and $\mu _{2}$ are the variation indices
associated with

\ \ \ \ \ \ \ large and small fractal scales, respectively.

Fig. 4- Box counting analysis for the signal used in Fig. 1.

\ \ \ \ \ \ \ \ \ \ $D_{B1}$\ and $D_{B2}$ are the box counting dimensions
associated

\ \ \ \ \ \ \ \ with large and small fractal scales, respectively.

Fig. 5- Hurst exponent $H_{1}$ (short-time correlations) as a function

\ \ \ \ \ \ \ \ of the standard deviation $\sigma $\ of the signal$.$

Fig. 6- Hurst exponent $H_{2}$ (long-time correlations) as a function

\ \ \ \ \ \ \ \ of the standard deviation \ $\sigma $\ of the signal.

Fig. 7- Box counting dimension $D_{B1}$\ (large fractal scale) as

\ \ \ \ \ \ \ a function of the standard \ deviation $\sigma $\ of the
signal.

Fig. 8- Box counting dimension $D_{B2}$ (small fractal scale) as

\ \ \ \ \ \ \ a function of the standard deviation $\sigma $\ of the signal.

Fig. 9- Variation index $\mu _{1}$\ (large fractal scale) as a\ \ function

\ \ \ \ \ \ \ \ of the standard deviation $\sigma $\ of the signal.

Fig. 10- Variation index $\mu _{2}$\ ( small fractal scale) as a function

\ \ \ \ \ \ \ \ \ \ of the standard deviation $\sigma $\ of the signal.

Fig. 11- DFA exponent $\alpha $\ as a function of the standard deviation $%
\sigma $

\ \ \ \ \ \ \ \ \ \ \ \ \ \ of the signal.

Fig. 12- Hurst exponent $H_{1}($short-time correlations) \ as a

\ \ \ \ \ \ \ \ \ function of the logarithm of the amplitude $A_{H1}.$

Fig. 13- Hurst exponent $H_{2}$\ (long-time correlations) as a

\ \ \ \ \ \ \ \ \ function of the logarithm of the amplitude $A_{H2}.$

Fig. 14- Box counting dimension $D_{B1}$(large fractal scale)\

\ \ \ \ \ \ \ \ \ as a function of the logarithm of the amplitude $%
A_{D_{B1}}.$

Fig. 15- Box counting dimension $D_{B2}$\ (small fractal scale)

\ \ \ \ \ \ \ \ \ \ as a function of the logarithm of the amplitude $%
A_{D_{B1}}.$

Fig. 16- Variation index $\mu _{1}$\ (large fractal scale) as a \ function

\ \ \ \ \ \ \ \ \ \ of the logarithm of the amplitude $A_{\mu 1}.$

Fig. 17- Variation index $\mu _{2}$\ (small fractal scale) as a\ \ function

\ \ \ \ \ \ \ \ \ of the logarithm of the amplitude $A_{\mu 2}.$

Fig. 18 - Tree summarizing all possible solutions for the discrimination

\ \ \ \ \ \ \ \ \ \ \ \ \ of the various failure modes, which correspond to
different paths

\ \ \ \ \ \ \ \ \ \ \ \ \ on the complete graph. The modes are shown on the
branches, and

\ \ \ \ \ \ \ \ \ \ \ \ \ the diagrams discriminating the modes are attached
to the nodes.


\begin{figure}
\begin{center}
\includegraphics[width=18cm]{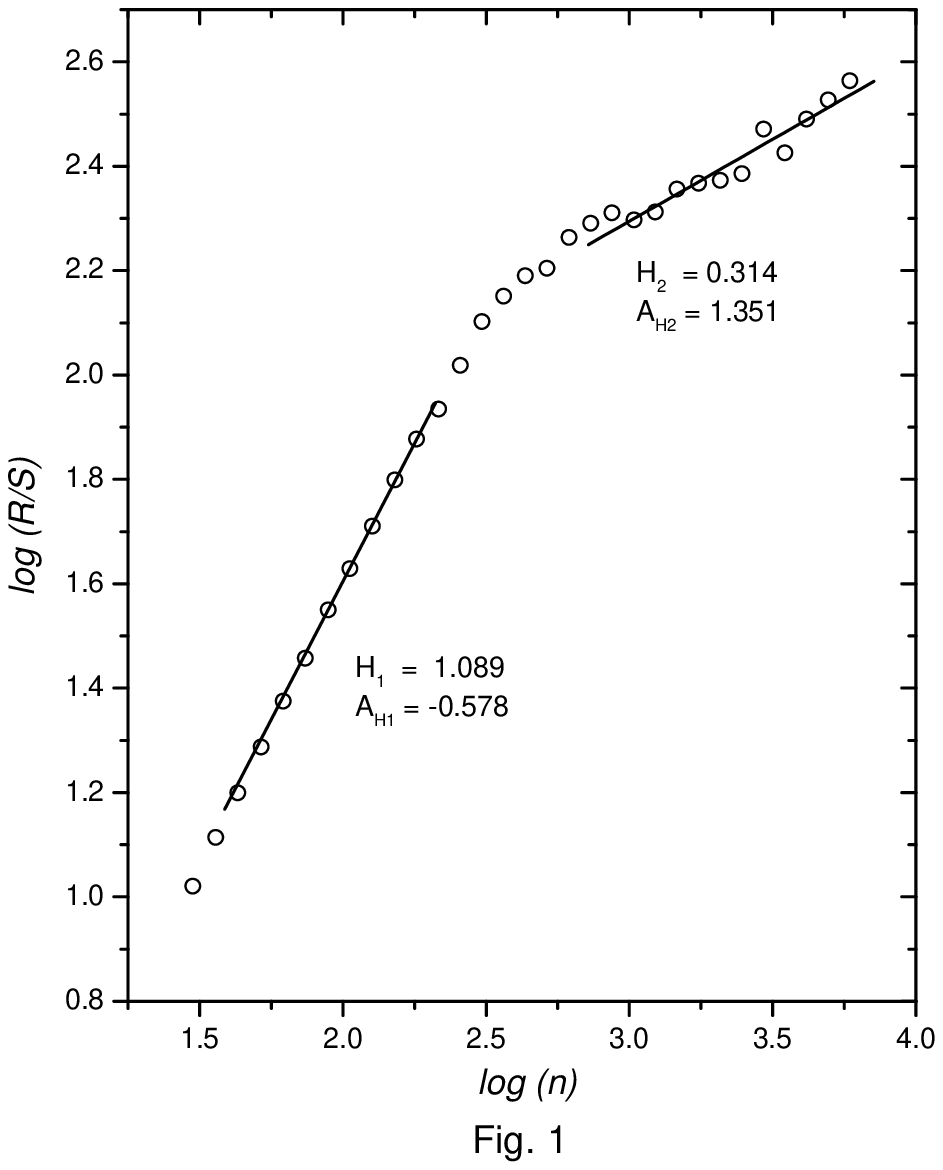}
\end{center}
\end{figure}

\begin{figure}
\begin{center}
\includegraphics[width=18cm]{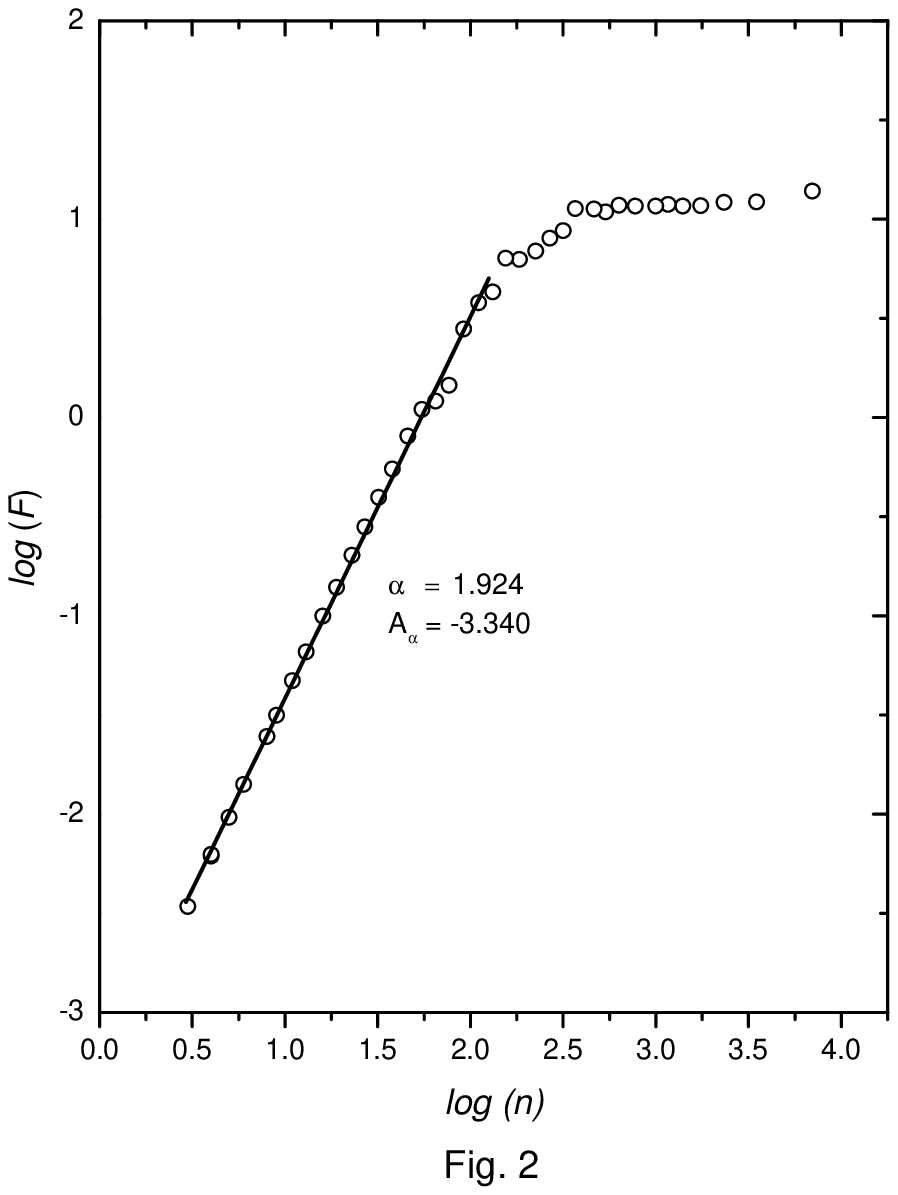}
\end{center}
\end{figure}

\begin{figure}
\begin{center}
\includegraphics[width=18cm]{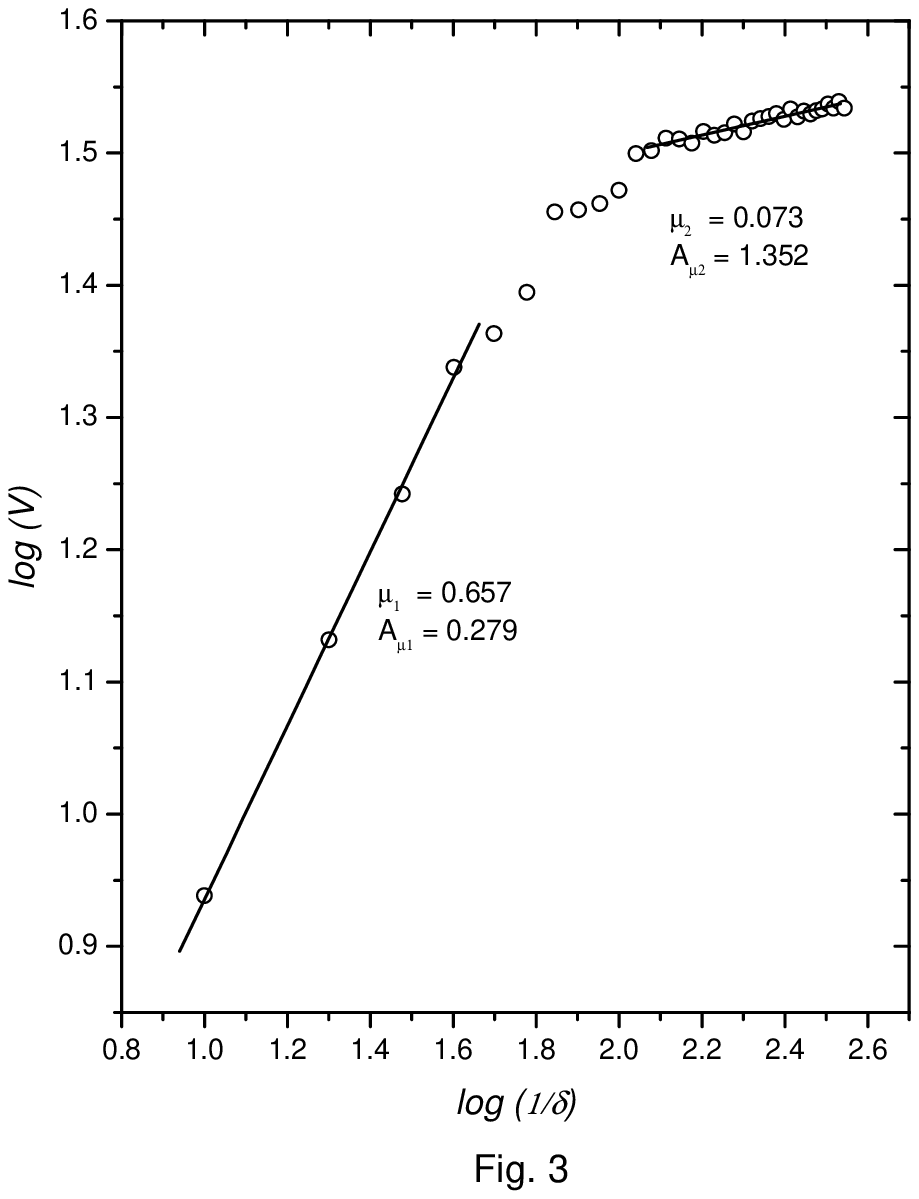}
\end{center}
\end{figure}

\begin{figure}
\begin{center}
\includegraphics[width=18cm]{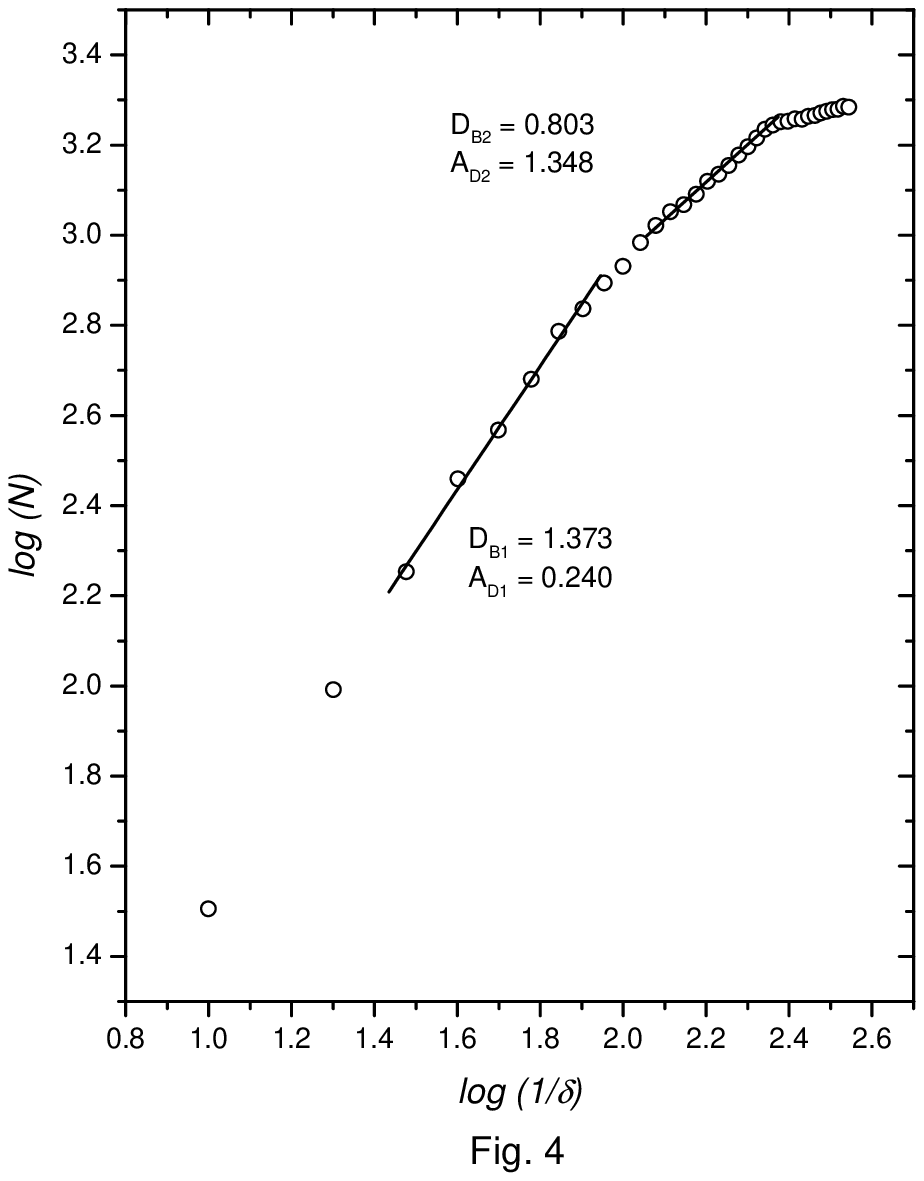}
\end{center}
\end{figure}

\begin{figure}
\begin{center}
\includegraphics[width=18cm]{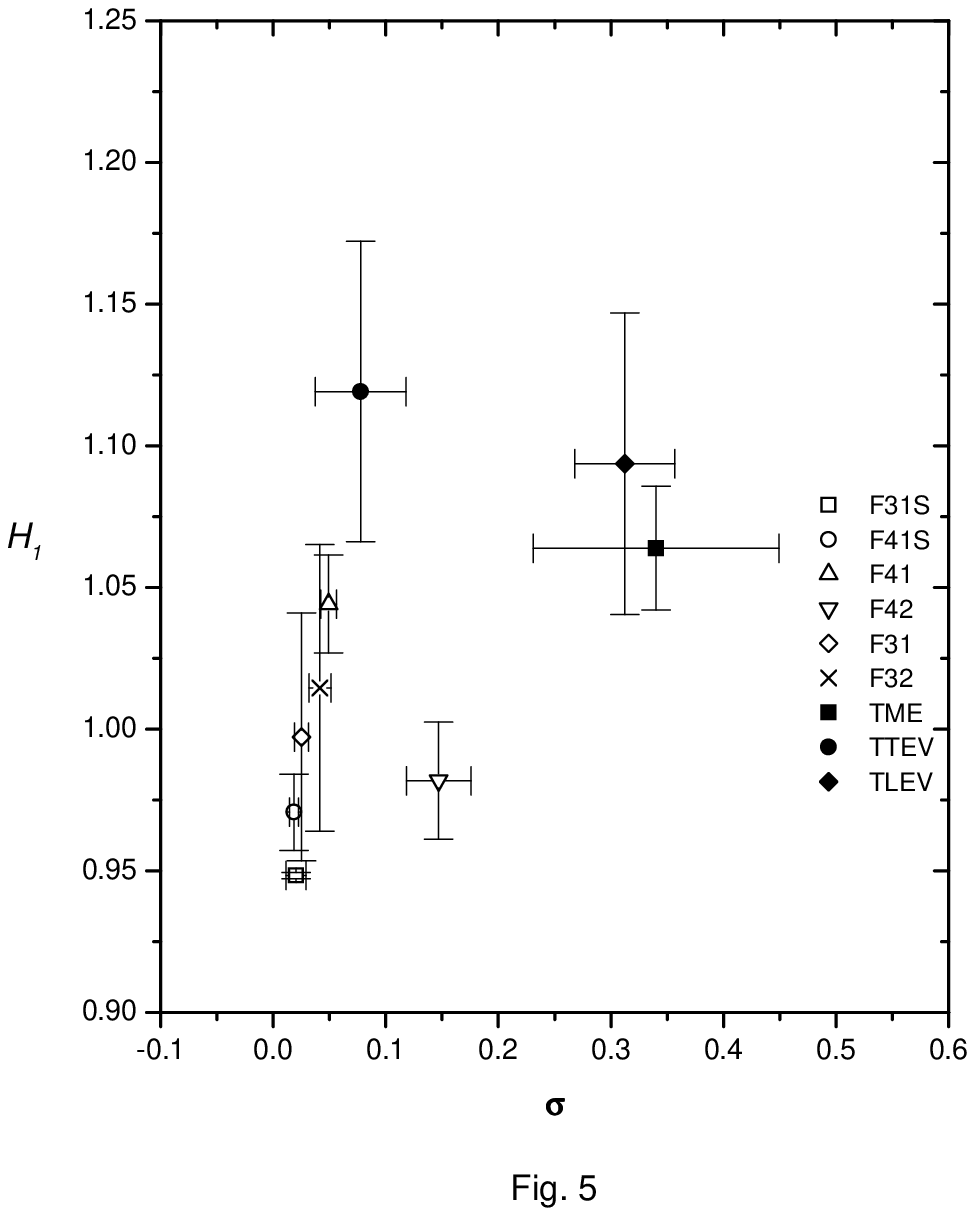}
\end{center}
\end{figure}

\begin{figure}
\begin{center}
\includegraphics[width=18cm]{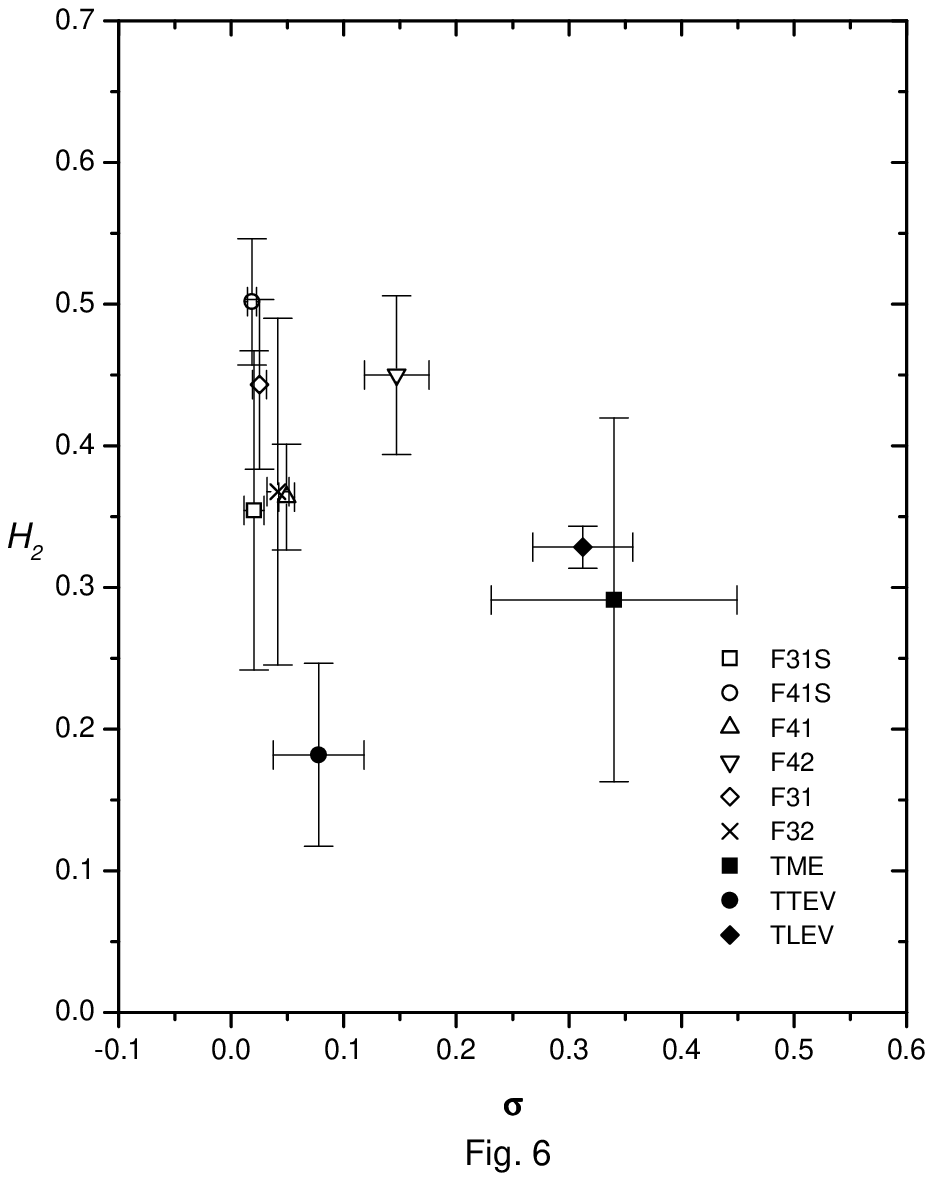}
\end{center}
\end{figure}

\begin{figure}
\begin{center}
\includegraphics[width=18cm]{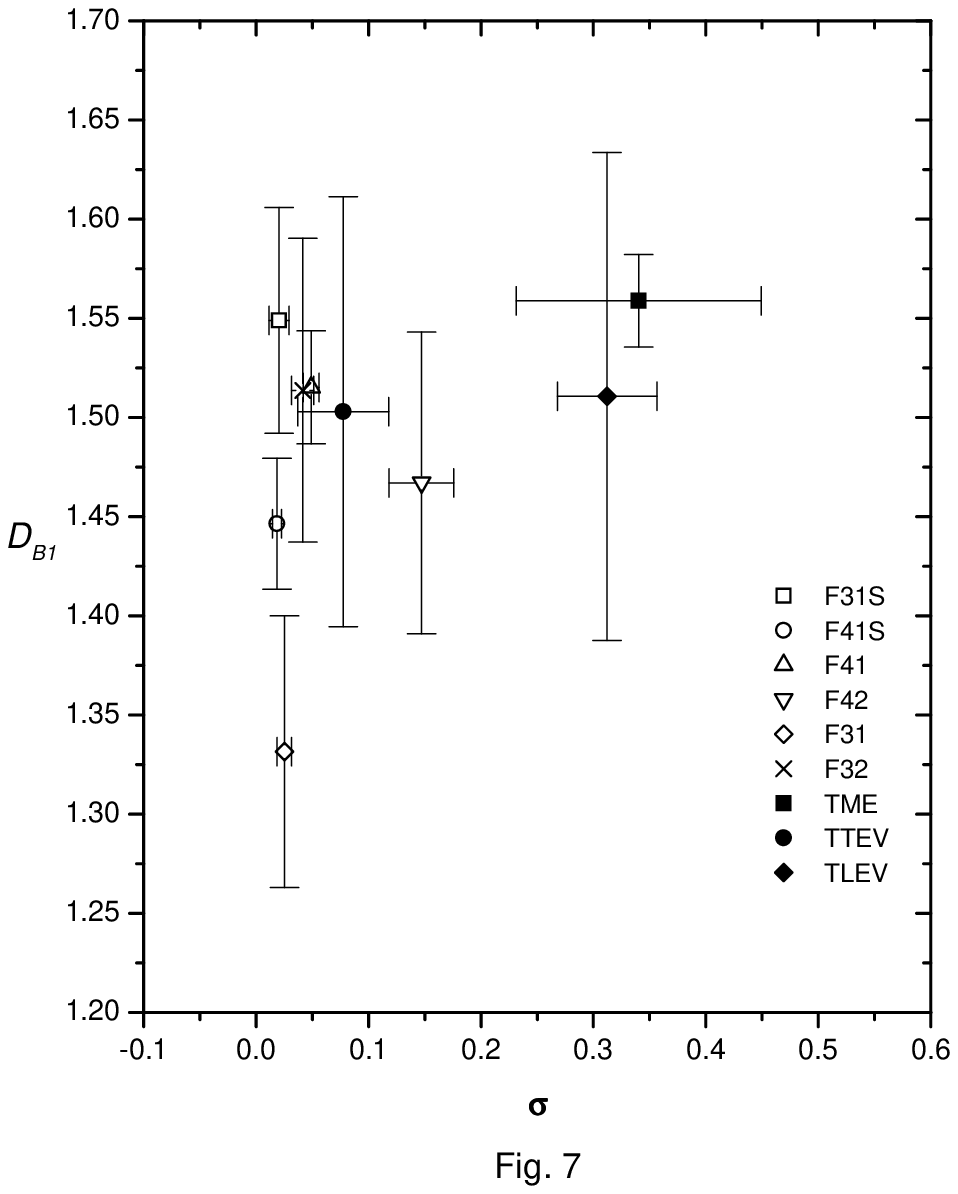}
\end{center}
\end{figure}

\begin{figure}
\begin{center}
\includegraphics[width=18cm]{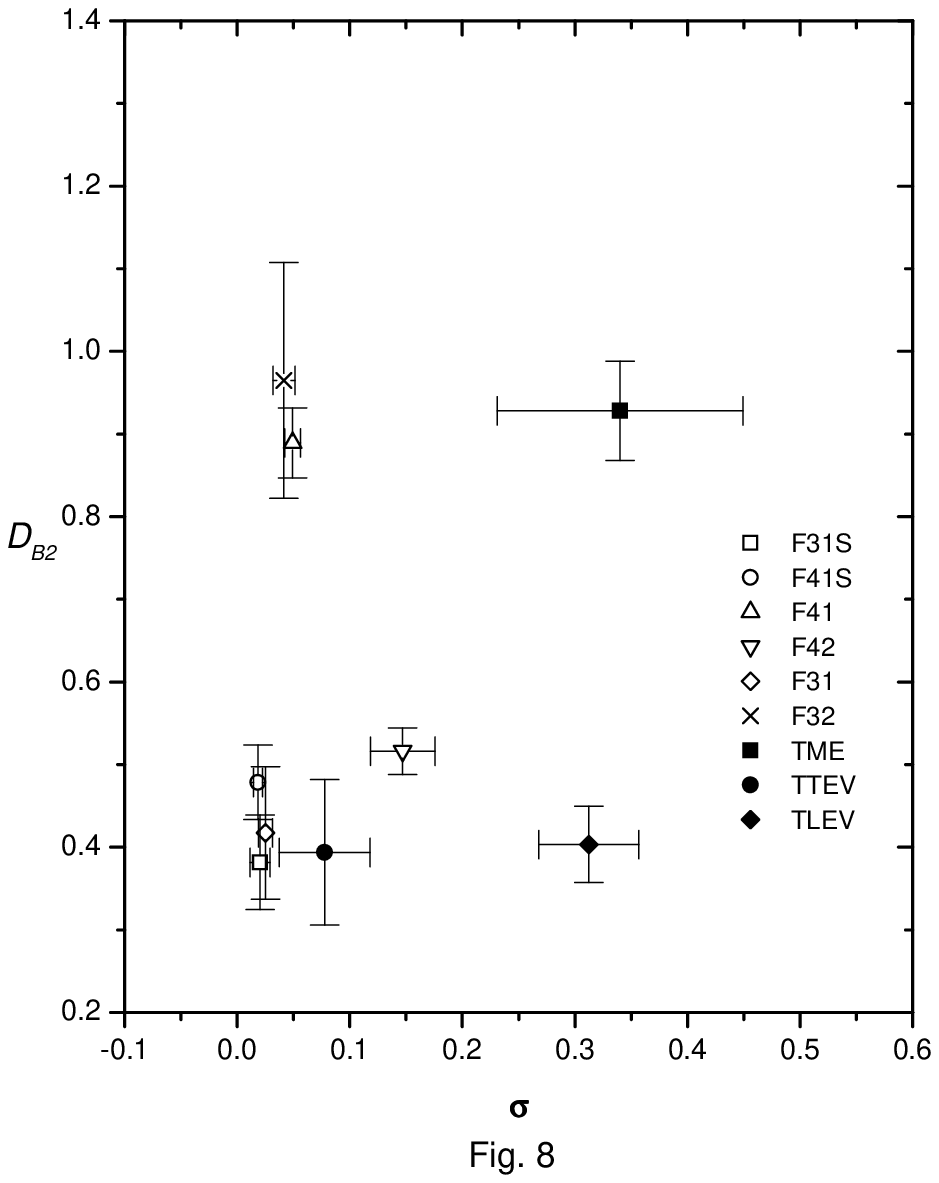}
\end{center}
\end{figure}

\begin{figure}
\begin{center}
\includegraphics[width=18cm]{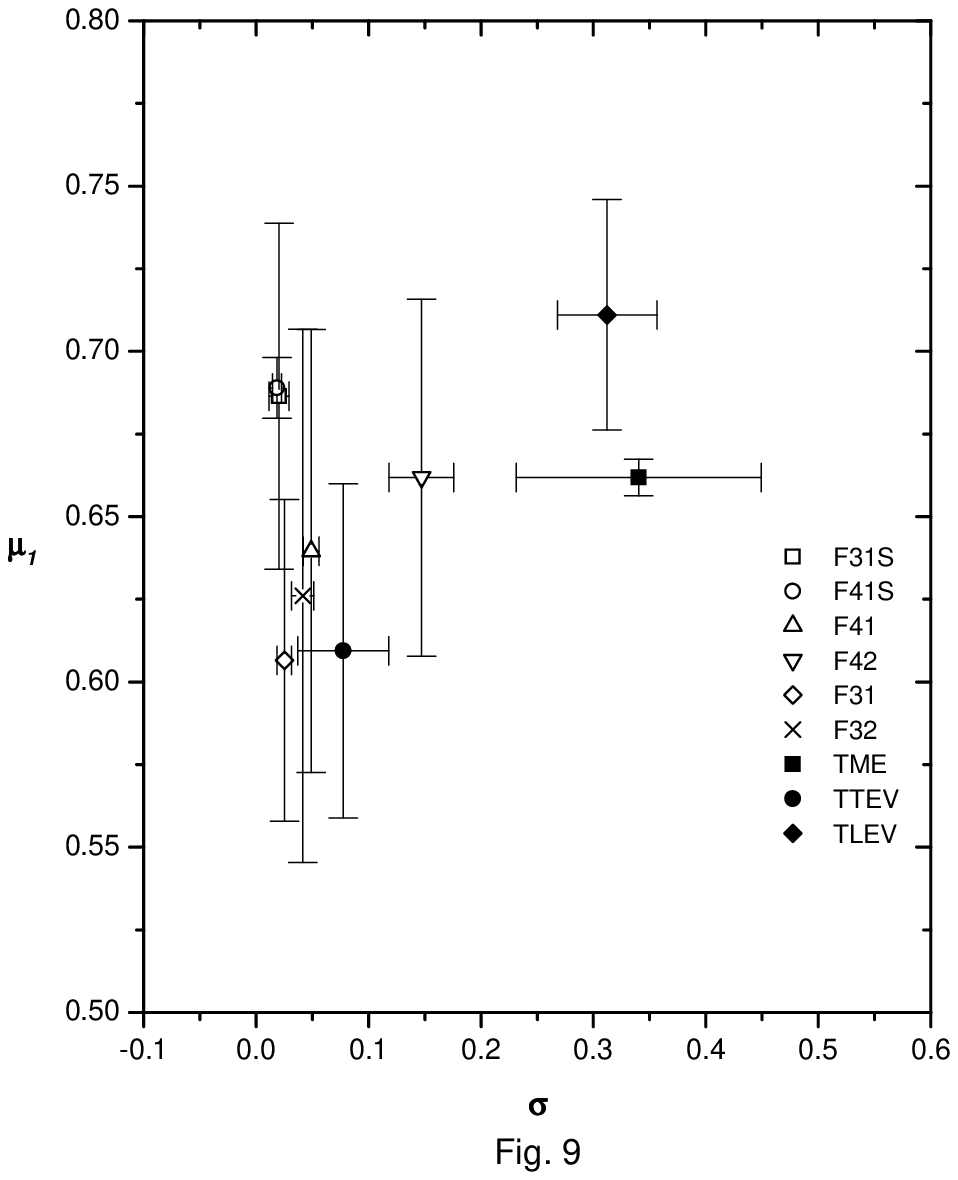}
\end{center}
\end{figure}

\begin{figure}
\begin{center}
\includegraphics[width=18cm]{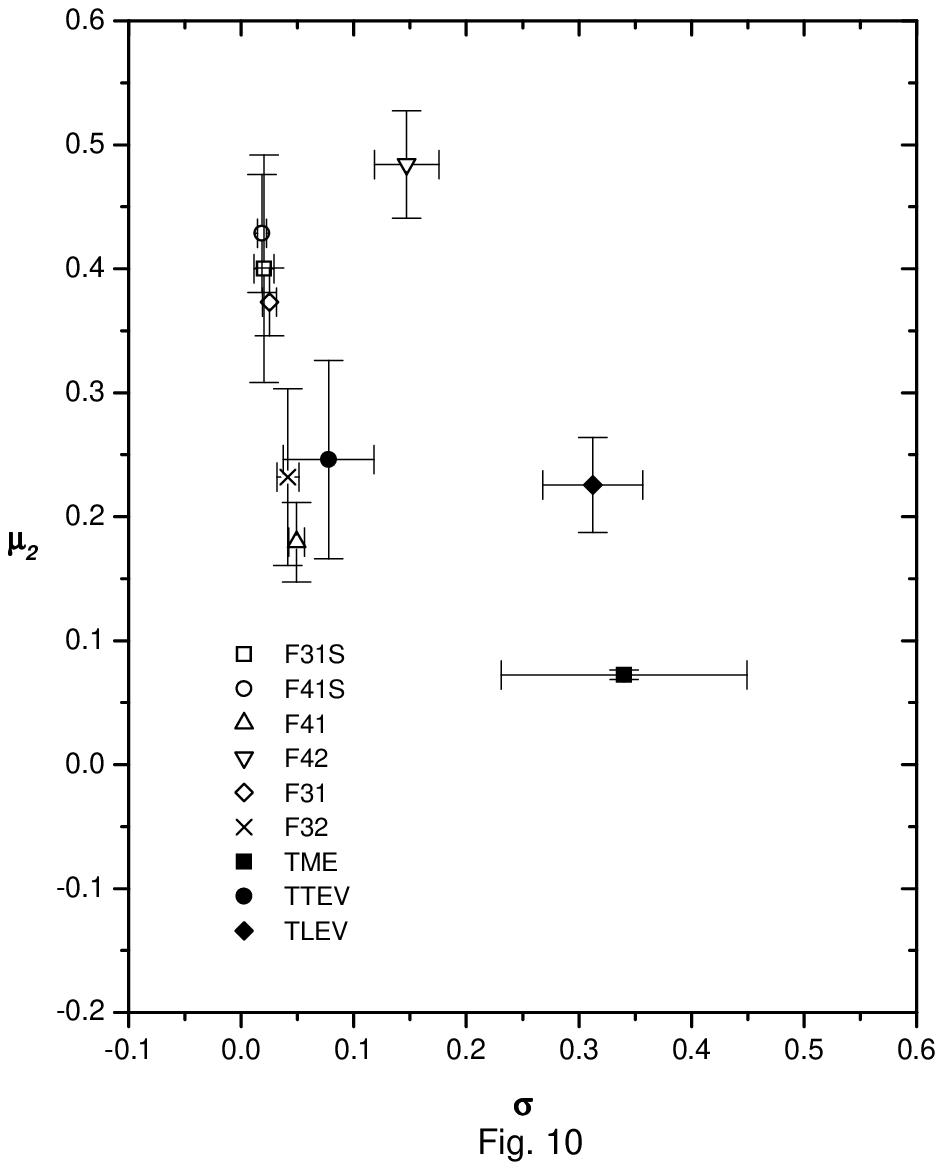}
\end{center}
\end{figure}

\begin{figure}
\begin{center}
\includegraphics[width=18cm]{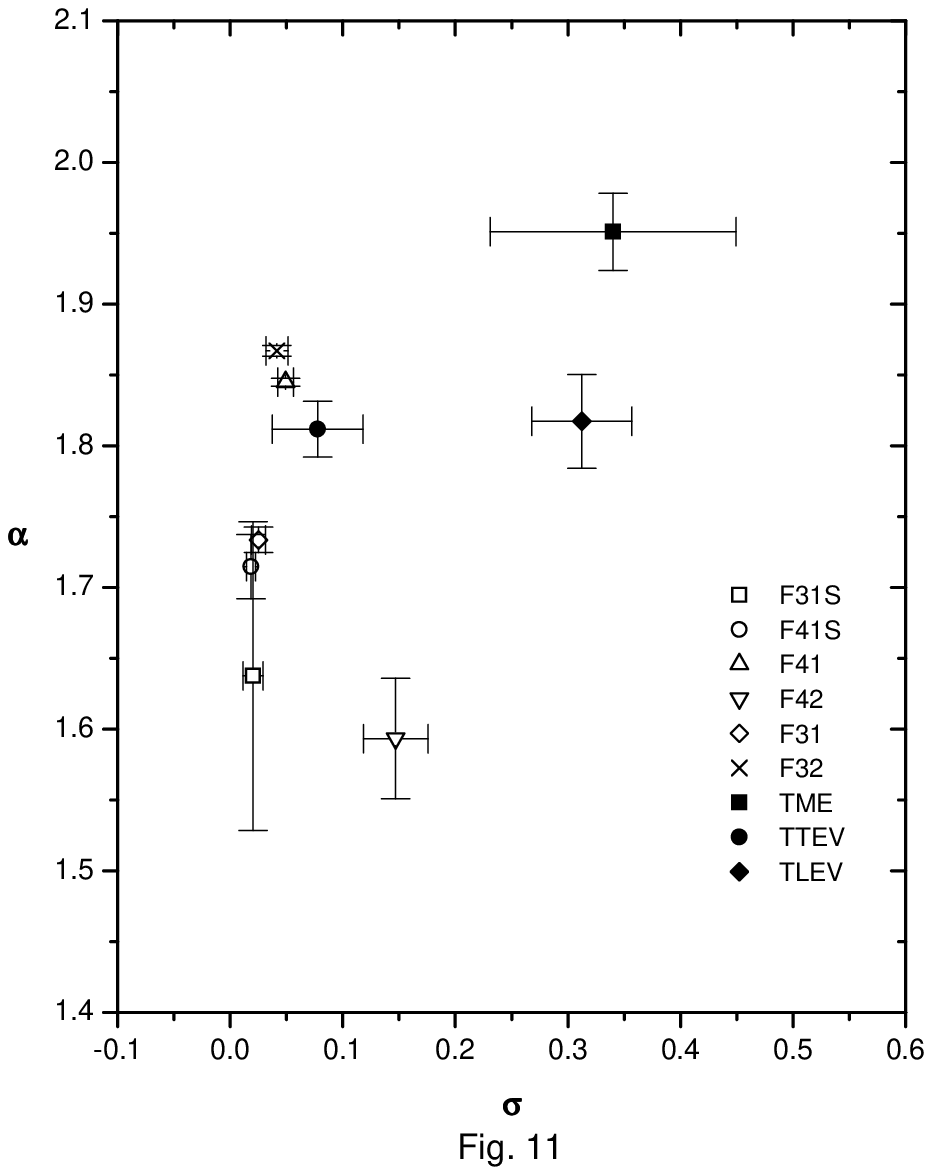}
\end{center}
\end{figure}

\begin{figure}
\begin{center}
\includegraphics[width=18cm]{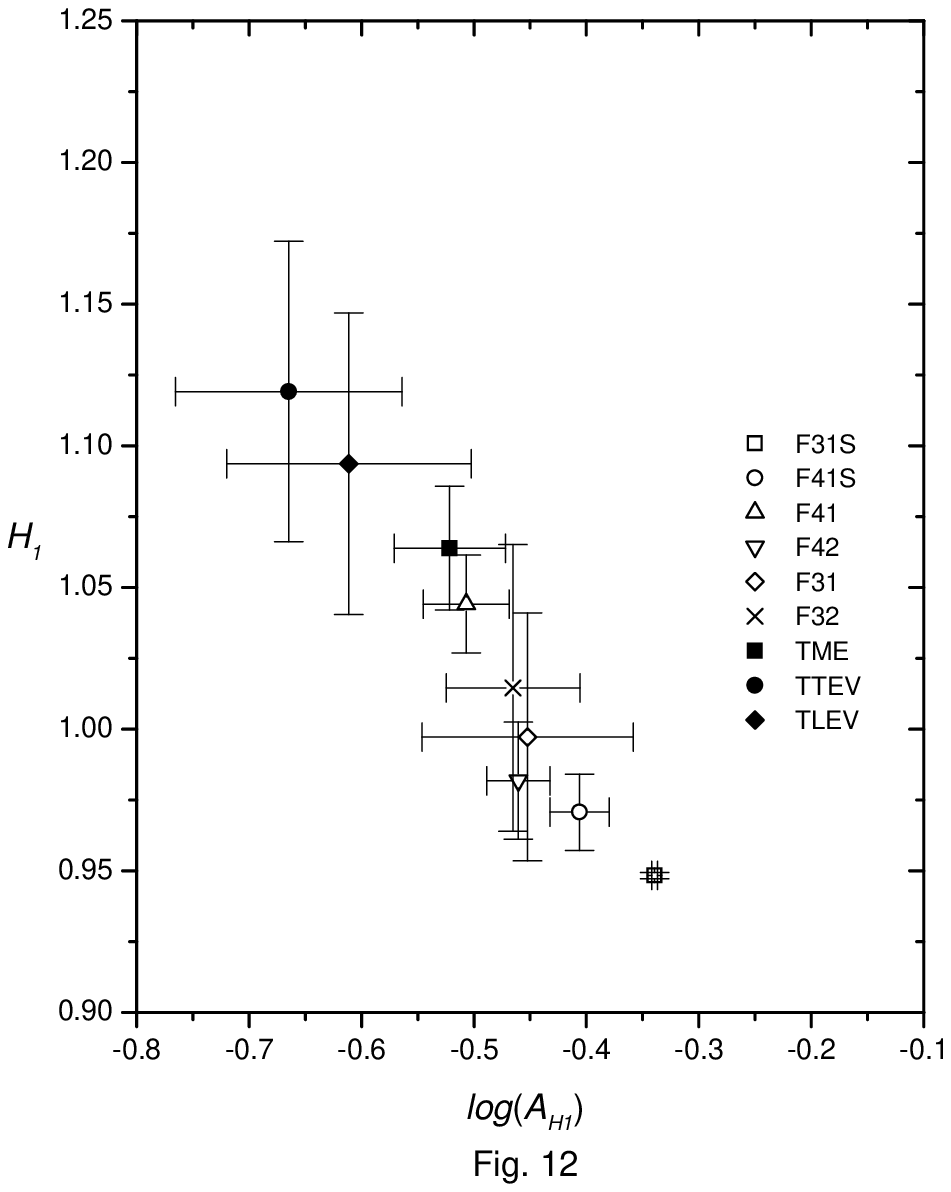}
\end{center}
\end{figure}

\begin{figure}
\begin{center}
\includegraphics[width=18cm]{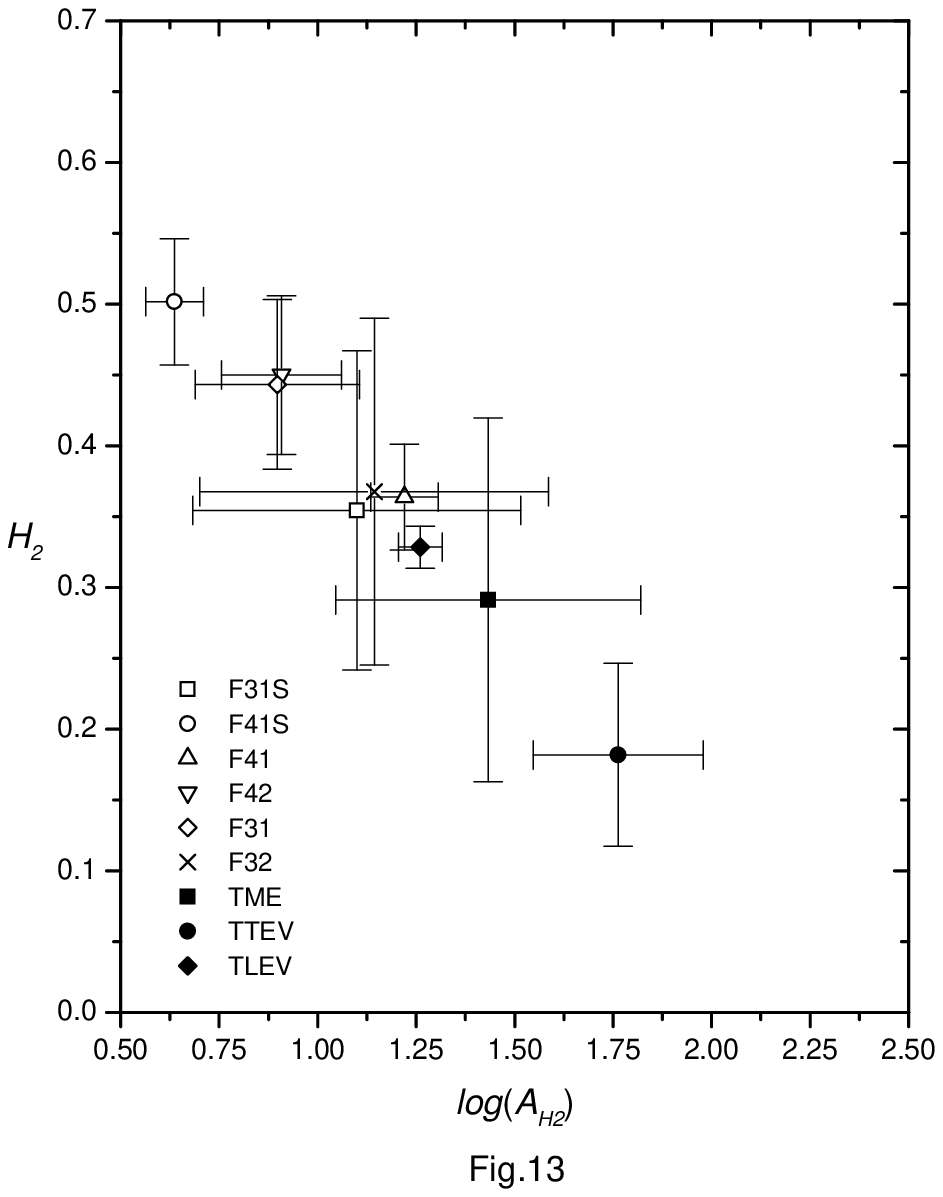}
\end{center}
\end{figure}

\begin{figure}
\begin{center}
\includegraphics[width=18cm]{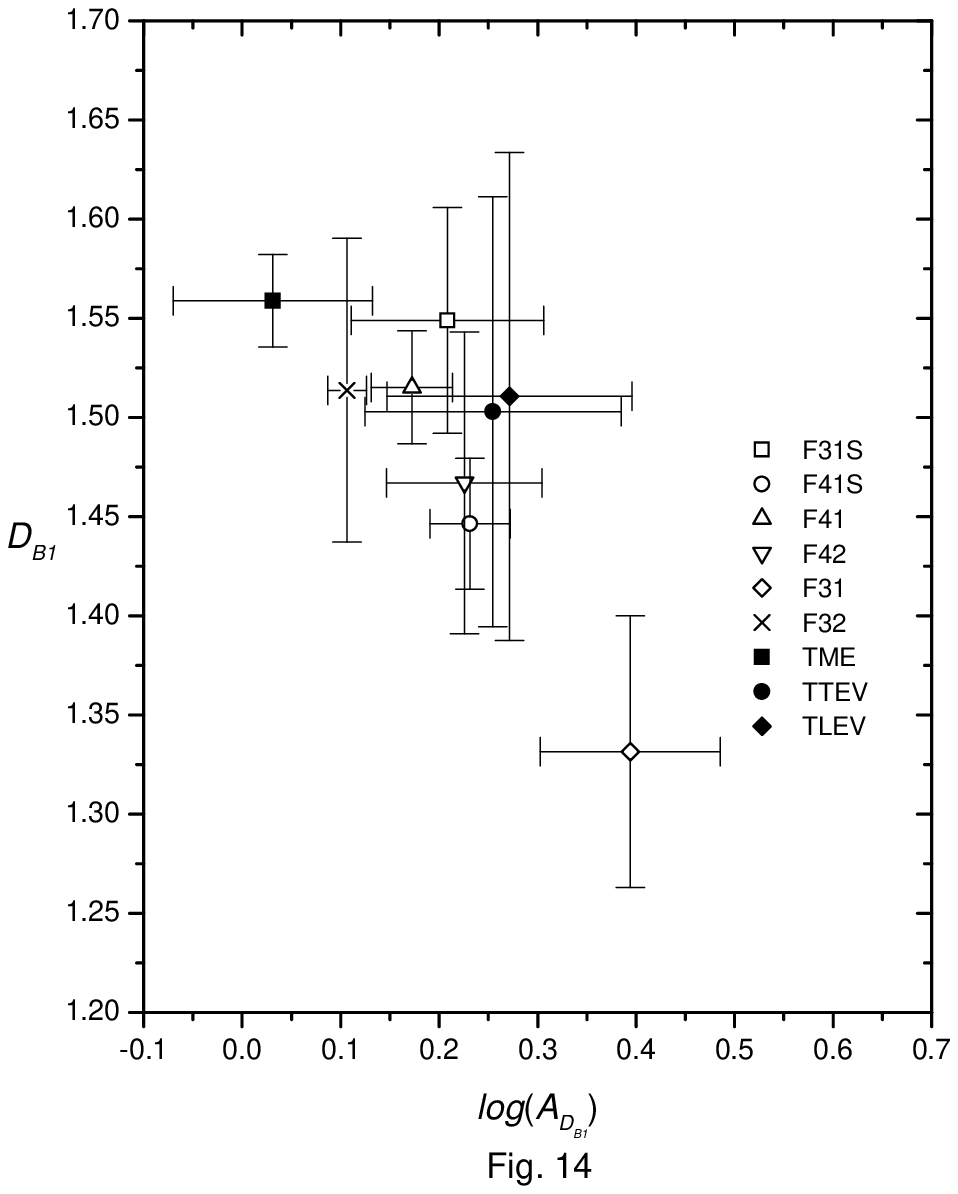}
\end{center}
\end{figure}

\begin{figure}
\begin{center}
\includegraphics[width=18cm]{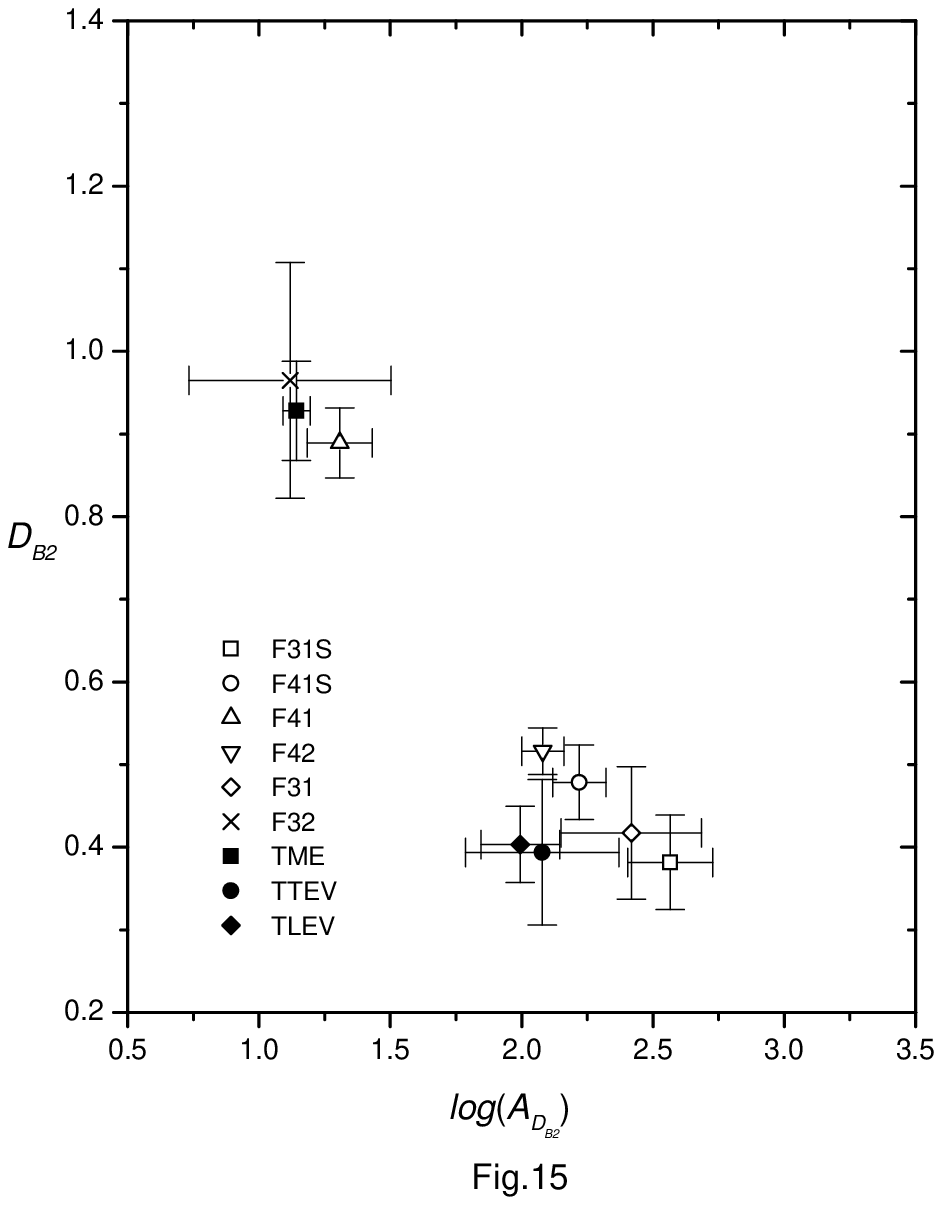}
\end{center}
\end{figure}

\begin{figure}
\begin{center}
\includegraphics[width=18cm]{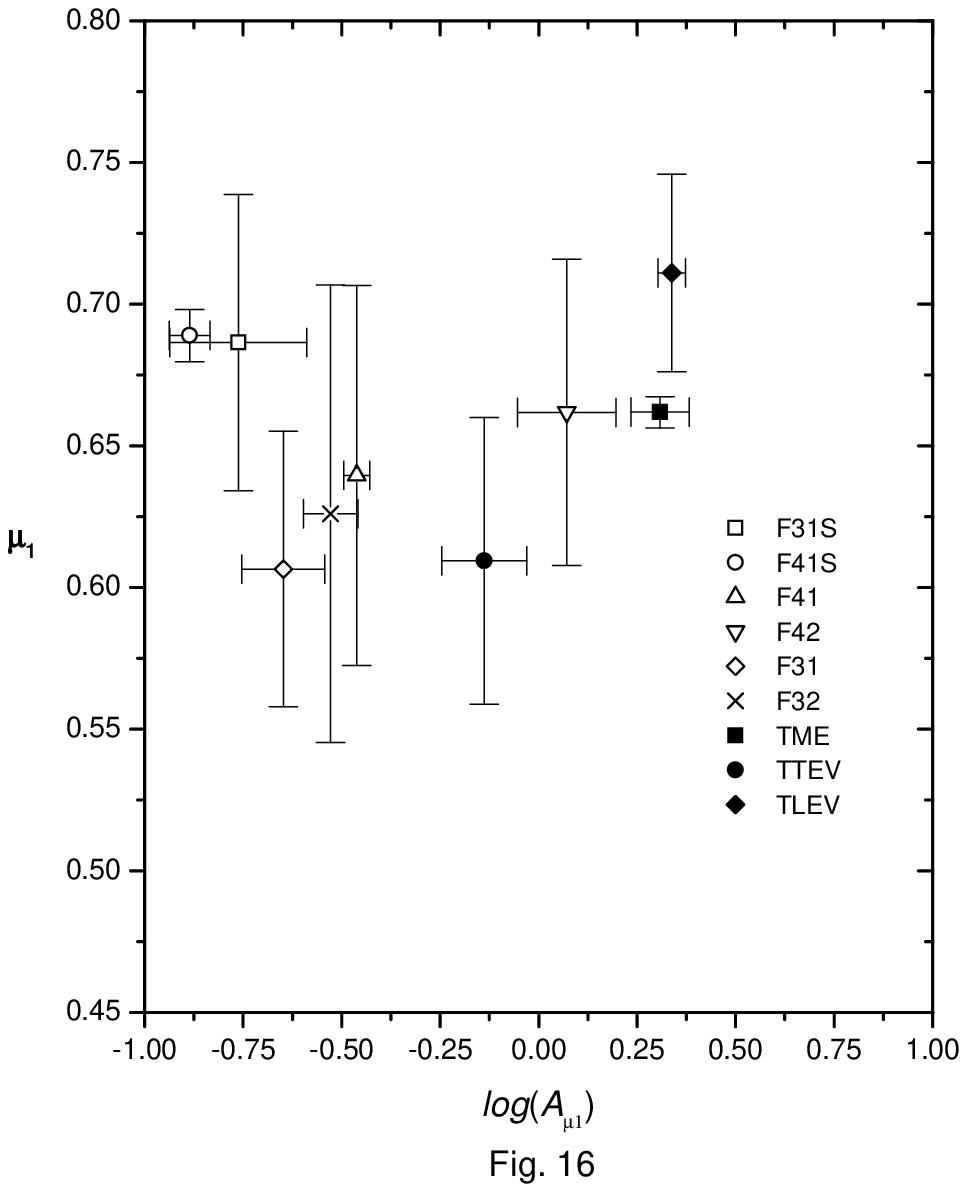}
\end{center}
\end{figure}

\begin{figure}
\begin{center}
\includegraphics[width=18cm]{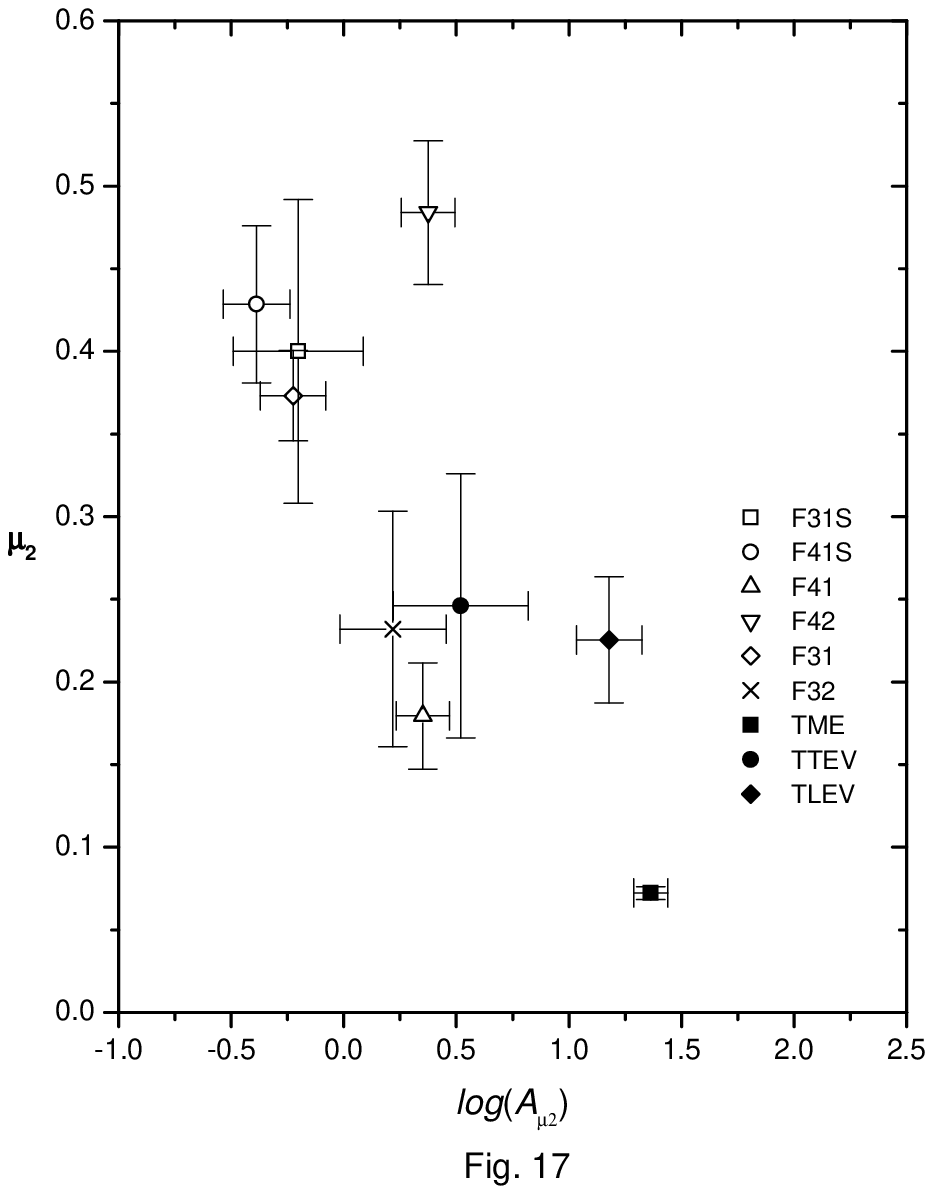}
\end{center}
\end{figure}

\begin{figure}
\begin{center}
\includegraphics[width=15cm]{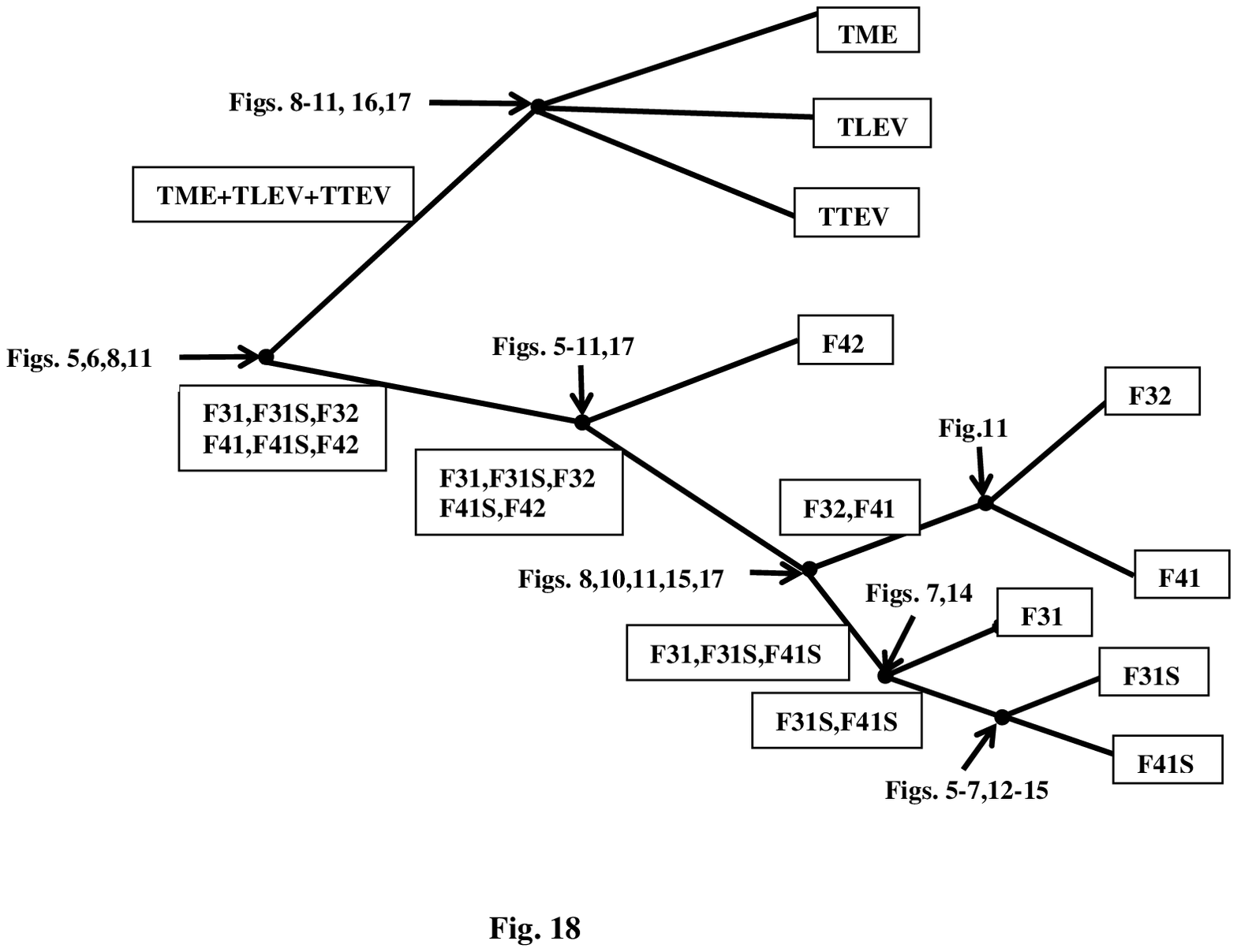}
\end{center}
\end{figure}

\end{document}